\documentstyle[preprint,version2,aps]{revtex}  % use this one
\begin{document}
\draft
\preprint{hep-ph/9510324}
\preprint{OCIP/C 95-10}
\preprint{April 1995}
\begin{title}
MEASUREMENT OF THE $WW\gamma$ and $WWZ$ COUPLINGS \\
 AT LEP200: THE BENEFITS OF HIGHER ENERGY?
\end{title}
\author{Mikul\'{a}\v{s} Gintner and Stephen Godfrey}
\begin{instit}
Ottawa-Carleton Institute for Physics \\
Department of Physics, Carleton University, Ottawa CANADA, K1S 5B6
\end{instit}
\begin{abstract}
We performed a detailed analysis of the process
$e^+e^-\to \ell \nu q\bar{q}'$ to determine its sensitivity
to anomalous trilinear gauge boson couplings of the $WW\gamma$
and $WWZ$ vertices and how the sensitivity varies with energy and
integrated luminosity.  We included all tree level Feynman diagrams that
contribute to this final state and used a
maximum likelihood analysis of a five dimensional differential
cross-section based on the $W$ and $W$ decay product angular distributions.
For constant luminosity, increasing $\sqrt{s}$ from 175~GeV to
192~GeV (220~GeV)
improves the measurement sensitivity by a factor of 1.5 to 2 (2 to 3)
depending on the parameter measured.
However, the lower luminosity expected at higher $\sqrt{s}$ will reduce
these improvements.  In any case, the sensitivities for
$\sqrt{s}$=175~GeV and L=500~pb$^{-1}$ of $\delta g_1^Z = \pm 0.22$,
$\delta \kappa_Z = \pm 0.20$, $\delta\kappa_\gamma=\pm 0.27$, $\delta
L_{9L}= \pm 55$, and $\delta L_{9R}= ^{+330}_{-230}$ are likely to be
at least an order of magnitude too big to see the effects of new physics.
\end{abstract}

\section{INTRODUCTION}
$e^+e^-$ colliders have contributed greatly to our knowledge of
electroweak interactions\cite{lep} and it is expected that
this tradition will continue in the future with the commissioning of
the CERN LEP-200 $e^+e^-$ collider.
One of the primary physics goals of LEP-200 is to
make precision measurements of  $W$ boson properties including precision
measurements of the $W$ mass, width, and $W$-boson couplings with
fermions and the photon and $Z^0$.

The latter measurements, that of
the trilinear gauge boson vertices (TGV's) provides a
stringent test of the gauge structure of the standard
model\cite{tgvreviews}.
The current measurement of these couplings are rather weak.
Using a popular parametrization of the CP conserving gauge
boson couplings, indirect measurements of TGV's via radiative corrections to
precision electroweak measurements\cite{burgess94,dawson95,hagiwara93}
give the following limits \cite{burgess94}:
$\delta g^Z_1 =-0.033\pm 0.031$, $\delta \kappa_\gamma =0.056\pm
0.056$, $\delta \kappa_Z =-0.0019\pm 0.044$,
$\lambda_\gamma=-0.036\pm 0.034 $, and $\lambda_Z =0.049\pm 0.045$.
However, there are ambiguities in these calculations so that these
limits are not particularly rigorous and it is necessary
to use unambiguous direct measurements for more reliable bounds.
The CDF and D0 collaborations at the Tevatron
$p\bar{p}$ collider at Fermilab, using the processes
$p\bar{p} \to W \gamma , \; WW, \; WZ$
have obtained the direct 95 \% C.L. limits of
$-1.6 <\delta\kappa_\gamma < 1.8$, $-0.6 < \lambda_\gamma < 0.6$,
$-8.6<\delta\kappa_Z < 9.0$, and $-1.7 < \lambda_Z < 1.7$
\cite{tevatron}.
These measurements are still quite weak but it
is expected that they will improve as the
luminosity of the Tevatron increases.

At LEP200 the gauge boson self-interactions
will be measured in $W$-pair production,
%% FOLLOWING LINE CANNOT BE BROKEN BEFORE 80 CHAR
\cite{tgvreviews,hagiwara87,lep200,gintner95,sekulin,papa94,berends95,couture92}.
One of the most useful of the $e^+e^-\to W^+W^-$ channels
is $e^+e^- \to \ell \nu q \bar{q}'$.
With only one unobserved neutrino this channel has the advantage
that it can be fully reconstructed
using the constraint of the initial beam energies
without problems discriminating the $W^+$ and $W^-$
and the QCD backgrounds that plague the fully hadronic
decay modes and offers much higher statistics than the fully
leptonic modes\cite{couture92}.
As a result of its importance there have been
numerous studies of this process including electroweak
radiative corrections to these reactions and the important
question of initial state radiation and the sensitivity of these
processes to anomalous $WW\gamma$ and $WWZ^0$ gauge boson couplings (TGV's)
%% FOLLOWING LINE CANNOT BE BROKEN BEFORE 80 CHAR
\cite{tgvreviews,hagiwara87,lep200,gintner95,sekulin,papa94,berends95,beenakker94,barklow92,fourfermions,berends94b,beenakker91,fleischer94,emrc}.

An important question for LEP200 is the sensitivity of the TGV
measurements to the centre of mass energy and
luminosity\cite{altarelli}.
In this letter we study the sensitivity of the four fermion final state
$e^+e^- \to \ell \nu_\ell q \bar{q}'$ where $\ell$ is either $e^\pm$
or $\mu^\pm$ and $q\bar{q}'$ can be either $(ud)$ or $(cs)$ to TGV's
for the centre of mass energies appropriate to
LEP200, $\sqrt{s}=175$, 192, 205, and 220~GeV, for various integrated
luminosities.  Our goal is to determine how
the measurements will be affected by changing the center of mass
energy and luminosity.

This issue has been addressed in a number of papers.  The classic
paper by Hagiwara, Peccei, Zeppenfeld, and Hikasa, \cite{hagiwara87}
examined the sensitivity of
anomalous TGV's to the process $e^+e^- \to W^+ W^-$ and how the sensitivity
varied with centre of mass energies relevant to LEP200.  Although
this paper did point out the importance of separating longitudinally
polarized $W$'s from transversely polarized $W$'s the analysis was
restricted to specific angular distributions and specific $W$ boson
polarizations and varied only one parameter at a time.
In addition it did not include the contributions
from the so-called background contributions; other tree level
diagrams that contribute to the same four fermion final state.

The information about the outgoing $W$ polarizations can be taken into
account by using the angular distributions of the $W$ boson decay
products.  Sekulin\cite{sekulin}
and Aihara {\it et al} \cite{tgvreviews} included this information
by using a binned maximum log likelihood
fit to a five dimensional differential cross section with respect to
the $W$ scattering angle and the polar and azimuthal decay angles of
the $W^+$ and $W^-$ bosons.  Both of these analysis looked at different
center of mass energies relevant to LEP200 but neither included the
background contributions.  The analysis of Sekulin assumed the
narrow width approximation for the $W$ decays.  The analysis by
Aihara {\it et al} was more sophisticated and included
initial state radiation, detector smearing, and various kinematic
cuts introduced to reduce backgrounds.  Aihara {\it et al} also assumed
relations among the parameters which in the language of the
non-linearly realized Chiral Lagrangian takes
$L_{9L}=L_{9R}$.  Their general conclusion that the sensitivity
to the TGV parameters increased by a factor of 1.5 going from
176~GeV to 190~GeV is consistent with what we find.

The recent studies by Berends and van Sighem \cite{berends95} and by
Papadopoulos \cite{papa94} are the closest in spirit to ours.  Both of these
studies included full tree level background processes and finite
width effects.  Berends and van Sighem also included initial state
radiation but did not consider the variations with center of mass
energy and did not quantify the sensitivities to TGV's.
Papadopoulos looked at a range of centre of mass energies but
restricted his study to
specific angular distributions and varied only one parameter at a time.

In this letter we tie all the various pieces of
previous analysis together; we include all tree level background
processes and finite width effects, we perform a binned log
likelihood fit to a five dimensional differential cross section, we
varied the different parameters simultaneously so that correlations
between them might show up, and we varied both the center of mass
energy and luminosities relevant to LEP200.

\section{The Effective Lagrangian}

We used two common parametrizations of the TGV's.
The first approach describes the $WWV$ verticies using the most general
parametrization possible that respects Lorentz invariance,
electromagnetic gauge invariance and $CP$ invariance
\cite{hagiwara87,gaemers79}. This approach has become the standard
parametrization used in phenomenology making the comparison
of the sensitivity of different measurements to the TGV's straightforward.
We do not consider CP violating operators in this paper as they  are tightly
constrained by measurement of the neutron
electron dipole moment which constrains the two CP violating parameters to
$|\tilde{\kappa}|, |\tilde{\lambda}|< {\cal O} (10^{-4})$ \cite{cp}.
With these constraints the $WW\gamma$ and $WWZ$ vertices have five
free independent parameters,
$g_1^Z$, $\kappa_\gamma$, $\kappa_Z$, $\lambda_\gamma$ and $\lambda_Z$ and
is given by \cite{hagiwara87,gaemers79}:
\begin{equation}
{\cal L}_{WWV} =  - ig_V \left\{ { g_1^V (W^+_{\mu\nu}W^{-\mu} -
W^{+\mu}  W_{\mu\nu} ) V^\nu
+ \kappa_V W^\dagger_\mu W_\nu V^{\mu\nu}
- {{\lambda_V}\over{M_W^2}} W^+_{\lambda\mu}W^{-\mu}_\nu V^{\nu\lambda}
}\right\}
\end{equation}
where the subscript $V$ denotes either a photon or a $Z^0$,
$V^\mu$ and $W^\mu$ represents  the photon or $Z^0$ and $W^-$
fields respectively,
$W_{\mu\nu}=\partial_\mu W_\nu-\partial_\nu W_\mu$ and
$V_{\mu\nu}=\partial_\mu V_\nu-\partial_\nu V_\mu$
and $M_W$ is the $W$ boson mass.  ($g_1^\gamma$ is constrained by
electromagnetic gauge invariance to be equal to 1.)
The first two terms correspond to dimension 4 operators and the
third term corresponds to a dimension 6 operator.  The mass in the
denominator of the dimension 6 term would correspond to the scale of
new physics, typically of order 1~TeV.  However, it has become the
convention to use the mass of the $W$ boson so that the $W$ magnetic
dipole and electric quadrupole can be written in a form similar to
that of the muon.  Nevertheless, one expects the dimension 6
operator to be suppressed with respect to the dimension 4 operators by
a factor of $M_W^2/(\Lambda=1\;\hbox{TeV})^2 \simeq 10^{-2}$.
Higher dimension operators correspond to momentum
dependence\cite{baur88} in the form factors which are not important
at LEP200 energies and will therefore not be considered.
At tree level the standard model requires $g_1^Z=\kappa_V=1$ and
$\lambda_V=0$.
Typically, radiative corrections from heavy particles will change
$\kappa_V$ by about 0.015 and $\lambda_V$ by about 0.0025 \cite{smloops}.
Because anomalous values of the $\lambda_V$ are expected to be
suppressed relative to those of $g_1^Z$ and $\kappa_V$ it is
extremely unlikely that interesting constraints can be placed on
them at LEP200 so we will not consider them further.

The second commonly used parametrization  is
the Chiral Lagrangian approach\cite{bagger93,nonlinear}.
A custodial $SU(2)$ is assumed which is supported to high accuracy by the
nearness of the $\rho$ parameter to 1. This approach assumes that the
theory has no light Higgs particles and the electroweak gauge bosons
interact strongly with each other above approximately 1~TeV.  This
can be described by a non-linear realization of the $SU(2)\times
U(1)$ symmetry in a chiral Lagrangian formalism leading to the
effective Lagrangian:
\begin{equation}
L= -i g {{L_{9L}}\over{16\pi^2}} Tr [W^{\mu\nu} D_\mu \Sigma
D_\nu \Sigma^\dagger  ]
-i g' {{L_{9R}}\over{16\pi^2}} Tr [B^{\mu\nu} D_\mu \Sigma^\dagger
D_\nu \Sigma ]
+ g g'{{L_{10}}\over{16\pi^2}} Tr [ \Sigma B^{\mu\nu} \Sigma^\dagger
W_{\mu\nu}]
\end{equation}
where $W_{\mu\nu}$ and $B_{\mu\nu}$ are the $SU(2)$ and $U(1)$ field
strength tensors given in terms of $W_\mu\equiv W_\mu^i \tau_i$ by
\begin{eqnarray}
W_{\mu\nu} & = & {1\over 2} (\partial _\mu W_\nu -\partial_\nu W_\mu
+{i\over 2} g[W_\mu , W_\nu ]) \nonumber\\
B_{\mu\nu} & = & {1\over 2} (\partial _\mu B_\nu -\partial_\nu
V B_\mu )\tau_3 ,
\end{eqnarray}
$\Sigma=\exp(iw^i\tau^i/v)$,  $v=246$~GeV, $w^i$ are the would-be
Goldstone bosons that give the $W$ and $Z$ their masses via the
Higgs mechanism, and the $SU(2)_L \times U(1)_Y$ covariant
derivative is given by
$D_\mu \Sigma = \partial_\mu \Sigma + {1\over 2} i g W_\mu^i \tau^i \Sigma
-{1\over 2} i g' B_\mu \Sigma \tau^3 $.  The Feynman rules are found
by going to the unitary gauge where $\Sigma=1$.  Note that the
coefficient $1/16\pi^2$ is often replaced with $v^2/\Lambda^2$.
$L_{10}$ contributes to the gauge boson self energies where it is tightly
constrained to $-1.1 \leq L_{10} \leq 1.5$ \cite{dawson95}
so we will not consider it
further.  New physics contributions are expected to result in values
of $L_{9L,9R}$ of order 1\cite{bagger93}.

The parameters from the two Lagrangians can be mapped onto each other
\cite{bagger93,nonlinear}:
\begin{displaymath}
\begin{array}{ll}
g_1^Z & = 1 +{{e^2}\over {32\pi^2 s^2 c^2}} (L_{9L} + {{2s^2
}\over{(c^2-s^2)}} L_{10}) \\
\kappa_z & = 1 + {{e^2}\over {32\pi^2 s^2 c^2}} (L_{9L}c^2 - L_{9R} s^2)
 +{{4s^2 c^2}\over {(c^2-s^2)}} L_{10} \\
\kappa_\gamma & = 1 + {1\over 32\pi^2} {e^2\over s^2} (L_{9L} +
L_{9R} - 2L_{10})
\end{array}
\end{displaymath}

\section{CALCULATIONAL APPROACH}

To study the process $e^+e^-\to \ell^\pm \nu q \bar{q}'$
we included all tree level diagrams to the four
fermion final states using helicity amplitude techniques\cite{calkul}.
The 10 diagrams contributing to the $e^+e^- \to \mu^\pm
\nu_\mu q \bar{q}'$ final state are shown in Fig. 1.
The gauge boson coupling we are studying is present in diagram (1a).
This, along with diagram (1b) are the diagrams responsible for real
$W$ production.
For the $e^\pm \nu_e q\bar{q}'$ final state the 10 diagrams shown in
Fig. 2 must also be included with those of Fig. 1 for a total
of 20 diagrams.
Diagram (2a) includes a TGV.  The diagrams with
t-channel photon exchange make large contributions to single $W$
production due to the pole in the photon propagator which can be
used to isolate the $WW\gamma$ vertex from the $WWZ$ vertex.

To evaluate the cross-sections and different distributions, we used the CALKUL
helicity amplitude technique \cite{calkul} to obtain expressions for the
matrix elements and
performed the phase space integration using  Monte Carlo
techniques \cite{monte}.
To obtain numerical results we used the values $\alpha=1/128$,
$\sin^2\theta=0.23$, $M_Z=91.187$ GeV, $\Gamma_Z=2.49$ GeV,
$M_W=80.22$ GeV, and
$\Gamma_W=2.08$ GeV.  In our results we included two generations of
quarks and took the quarks to be massless.
In order to take into account finite detector acceptance we require
that the lepton and quarks are at least
10 degrees away from the beam and have at least 10~GeV energy.

In principle we should include QED radiative corrections from soft photon
emission and the backgrounds due to a photon that is lost
down the beam pipe \cite{berends94b,emrc}.  These backgrounds are well
understood and detector dependent.  We assume the approach taken at
LEP, that these effects can best be taken into account by the
experimental collaborations.  In any case, although initial state radiation
must be taken into account their inclusion does not
substantially effect the bounds we obtain and therefore our conclusions.

Our primary interest here is to examine the sensitivity of TGV
measurements to $\sqrt{s}$
\cite{tgvreviews,hagiwara87,sekulin,papa94}
and to integrated luminosity.
Any disruption of the delicate gauge theory cancellations leads to
large changes to the standard model results.  The corrections for
$W_L$ production amplitudes can be enhanced by a factor of $(s/M_W^2)$
\cite{hagiwara87}.
Because it is the longitudinal $W$ production which is most
sensitive to anomalous couplings it is crucial to disentangle
the $W_L$ from the $W_T$ {\sl background}.    The most convenient
means of doing so makes use of the angular distributions of the $W$
decay products.  We define the 5 angles:
$\Theta$, the $W^-$
scattering angle with respect to the initial $e^+$ direction,
$\theta_{qq}$, the polar decay angle of the $q$ in the $W^-$ rest
frame using the $W^-$ direction as the quantization axis,
$\phi_{qq}$, the azimuthal decay angle of the $q$ in the $W^-$ rest
frame, and $\theta_{\ell\nu}$ and $\phi_{\ell\nu}$ are the analogous
angles for the lepton in the $W^+$ rest frame. These angles are shown in
Fig. 3. We define the azimuthal
angle as the angle between the normal to the reaction
plane, $n_1=p_e \times p_W$ and the plane defined by the $W$ decay
products, $n_2=p_q \times p_{\bar{q}}$.
The angular distribution in $\theta$ peaks about
$\cos\theta=0$ for longitudinally polarized $W$ bosons and at
forward or backward angles for transversely polarized bosons. In
addition the parity violation of the $W$ couplings distinguishes the
two polarization states adding to the effectiveness of the decay as
a polarimeter.   Thus, the angular distributions can be used to
extract information about the $W$ boson polarizations.

To use the information contained in these angular distributions
we performed a maximum likelihood analysis
based on the 5 angles described above\cite{gintner95,sekulin,barklow92}.
For the $q\bar{q}$
case there is an ambiguity since we cannot tell which hadronic jet
corresponds to the quark and which to the antiquark.  We therefore
include both possibilities in our analysis.
To impliment the maximum likelihood analysis we divided each
of $\Theta$, $\theta_{qq}$, $\phi_{qq}$, $\theta_{\ell\nu}$, and
$\phi_{\ell\nu}$ into four bins so that the entire phase space was
divided into $4^5 = 1024$ bins.
Given the cross-section of $\sim 1$~pb per mode
with this many bins some will not
be very populated with events so that it is more appropriate to use
Poisson statistics rather than Gaussian statistics.  This leads
naturally to the maximum likelihood method.  The change in the
log of the Likelihood function from the standard model expectation
is given by
\begin{equation}
\delta\ln {\cal L} = \sum [ -r_i +r_i \ln (r_i) +\mu_i -r_i \ln (\mu_i) ]
\end{equation}
where the sum extends over all the bins and $r_i$ and $\mu_i$ are
the predicted number of non-standard model and standard model
events in bin $i$ respectively, given by
\begin{equation}
r_i = L \int_{\Delta\Theta} \int_{\Delta\theta_{qq} }
\int_{\Delta\phi_{qq}} \int_{\Delta_{\ell\nu}}
\int_{\Delta\phi_{\ell\nu}} {{d^5\sigma}\over{d\cos\Theta
d\cos\theta_{qq} d\phi_{qq} d\cos\theta_{\ell\nu} d\phi_{\ell\nu} }}
d\cos\Theta \; d\cos\theta_{\ell\nu} \; d\phi_{\ell\nu} \;
d\cos\theta_{qq} \; d\phi_{qq}
\end{equation}
where $ L$ is the expected integrated luminosity.
The 68\% and 95\% confidence level bounds are given by the values of
anomalous couplings which give a change in $\ln{\cal L}$ of 0.5 and
2.0 respectively.

The results presented here are based solely on the statistical errors based
on the integrated luminosity we assume for the various cases.
To include the effects of systematic errors using the maximum likelihood
approach requires an unweighted Monte Carlo simulation through a
realistic detector.  Since we did not have the facilities to do this we
attempted a simplified estimate of systematic errors using
a $\chi^2$ analysis to make our estimates.
Assuming a systematic 5\% measurement error combined
in quadrature with the statistical error we found that the systematic
errors are neglible compared to the statistical errors for the
integrated luminosities anticipated at LEP200.  That this is so is
a consequence of the large number of bins resulting in
a small number of events per bin leading to a large statistical error.
Thus, it appears as if the systematic errors will be dominated by
statistical errors but clearly, a full detector Monte Carlo must be
performed to properly understand the situation.

\section{RESULTS AND DISCUSSION}

A thorough analysis of gauge boson couplings would allow all
parameters in the Lagrangian to vary simultaneously to take into
account cancellations (and correlations) among the various contributions.
This approach is  impractical, however, due to the large
amount of computer time that would be required to search the parameter
space.  Instead we show 2-dimensional C.L. contours for a selection
of parameter pairs to give a sense of the correlations.
We believe
that these contours are reasonably reliable as when
the other parameters at the edges of our 2-dimensional contours were
varied, there was little change in the sensitivities.
For the case
of the Chiral Lagrangian where the global SU(2) symmetry imposes
relations between the parameters and where we restrict ourselves to
dimension four operators the parameter space reduces to 2 dimensions.

For the results we present here we did not include
a cut on $M_{\ell\nu}$ or
$M_{q\bar{q}}$ as these cuts in general have virtually no effect on the
sensitivities except for the electron mode involving the $WW\gamma$
vertex where the effect is still quite small.
We calculated the sensitivities of anomalous couplings for
$\sqrt{s}=175,\; 192, \; 205$, and 220~GeV assuming the same
integrated luminosity of 500~pb$^{-1}$ for all cases for the
purposes of comparison.  We show the sensitivities that can be
obtained by combining the $e^+$, $e^-$, $\mu^+$, and $\mu^-$ modes
to improve the statistics.
The 95\% confidence limit contours for the $g^Z_1-\kappa_Z$,
$\kappa_\gamma -\kappa_Z$, and $L_{9L}-L_{9R}$ planes are shown in
Fig. 4.
The sensitivities of the couplings, varying one parameter at a time,
are summarized in Table I.

At threshold, anomalous couplings are quite sensitive to energy with
improvements in sensitivies going from $\sqrt{s}=175$~GeV to
$\sqrt{s}=192$~GeV ranging from about 1.7 for $\kappa_Z$ to $\sim 2$ for
$L_{9R}$. The corresponding changes going from
$\sqrt{s}=205$~GeV to
$\sqrt{s}=220$~GeV are 1.2 for $\kappa_Z$ and $\sim 1.4$ for $L_{9R}$.
This occurs even though the cross section only varies from 1.10~pb (1.15~pb) at
175~GeV
to its maximum value of 1.28~pb (1.31~pb) at 200~GeV for the $\mu$ $(e)$
mode. We included the angular cuts but not the energy cuts on the final
state fermions to obtain these numbers.

The sensitivity can be understood by examining the helicity
amplitudes in detail which are given by Hagiwara {\it et
al.} in Ref. \cite{hagiwara87}.  At threshold the $\nu$-exchange
diagram dominates so that at low energy the $\gamma$ and $Z$
diagrams are down by a factor of $\beta=\sqrt{1-4M_W^2/s}$
compared to the $\nu$-exchange diagram.  Thus, at threshold the
cancellation between the various diagrams are less important than
at higher energies.
Near threshold the contributions from anomalous couplings go roughly
like $\beta$ with a further enhancement of roughly
$\sqrt{s}/2M_W$ for each longitudinal $W$ boson in the final state.
The net results is that in the threshold region the reaction
$e^+e^-\to W^+W^-$ is not very sensitive to the 3-vector boson
coupling.

The improvement in sensitivities going to higher energies is
slightly misleading as we have assumed that the luminosities would
be the same in all cases.  In reality one expects that the
luminosities will be lower for the higher energies \cite{altarelli}.
The coupling measurements are limited by statistics which are in
turn related to the number of events so that the sensitivities are
inversely proportional to the square root of the integrated
luminosity; $\delta \propto L^{-1/2}$.  Reducing the luminosity from
500~pb$^{-1}$ to 300~pb$^{-1}$ decreases the sensitivity by a factor
of 1.3 which limits the usefulness of increasing the center of mass energy.

In any case, these limits are at the very least
an order of magnitude less sensitive than would be required to see
the effects of new physics through radiative corrections and are
comparable to the sensitivities that could be achieved at a high
luminosity Tevatron upgrade.  It is therefore unlikely, that new
physics will reveal itself at LEP200 through precision measurements
of the TGV's.

\section{CONCLUSIONS}

The primary purpose of this note was to examine the sensitivity of
anomalous gauge boson coupling measurements at LEP200 to changes in
the center of mass energy.  As $W$ pair production is relatively
insensitive to TGV's near threshold a modest increase in energy from
$\sqrt{s}=175$~GeV to 192~GeV could yield sizable improvements in
the measurements.  Subsequent increases in energy to 205~GeV and
220~GeV would not yield the same improvement.  However, this
statement must be tempered with the reality that increases in energy
would likely result in lower luminosities and statistics with the
corresponding measurement degradation.

\acknowledgments

The authors greatfully acknowledge the participation of Gilles
Couture in related work.
The authors benefitted greatly from many helpful conversations,
communications, and suggestions  during the course of this work
with Tim Barklow, Genevieve B\'elanger, Pat Kalyniak and Paul Madsen.
This research was supported in part by the Natural Sciences and
Engineering Research Council of Canada.

\figure{The Feynman diagrams contributing to the process
$e^+e^- \to \mu^+ \nu_\mu q \bar{q}'$.}

\figure{The Feynman diagrams that contribute to the process
$e^+e^- \to e^+ \nu_\mu q \bar{q}'$ in addition to those of fig 1.}

\figure{Angle definitions used in our analysis.  $\Theta$ is the $W$
scattering angle, $\theta_{qq}$ and $\theta_{\ell\nu}$ are the decay
angles in the $W$ rest frame and $\phi_{qq}$ and $\phi_{\ell\nu}$ are
the azimuthal angles, again the $W$ rest frames.}

\figure{95\% C.L. contours for sensitivity to anomalous couplings for
$\sqrt{s}=175,\; 192, \; 205,$ and 220~GeV.  In all cases the
contours are obtained from
combining all 4 lepton charge states for L=500~pb$^{-1}$.  The
contours correspond to increasing energy going from the outer
contour to the inner with the outermost
contour corresponding to $\sqrt{s}=175$~GeV and the innermost
corresponding to $\sqrt{s}=220$~GeV.}

\begin{table}
\caption{Sensitivities to anomalous couplings for the various
parameters varying one parameter at a time.  The values are obtained
by combining the four lepton modes ($e^-$, $e^+$, $\mu^-$, and
$\mu^+$) and two generations of light quarks ($ud$, $cs$).  The
results are  95\% confidence level limits for the given integrated
luminosities.}
\label{thetable}
\begin{tabular}{lllllll}
$\sqrt{s}$ & L& $g_1^Z$ & $\kappa_Z$ & $\kappa_\gamma$ & $L_{9L}$ & $L_{9R}$\\
(GeV) & (pb$^{-1}$) & & & & & \\
\tableline
175 & 500 & $\pm 0.22$ & $^{+0.19}_{-0.20}$ & $^{+0.27}_{-0.26}$
	& $\pm 55$ & $^{+330}_{-230}$ \\
175 & 300 & $^{+0.28}_{-0.29}$ & $\pm 0.25$ & $^{+0.36}_{-0.33}$
	& $\pm 70$ & $^{+440}_{-300}$ \\
192 & 500 & $\pm 0.15$ & $\pm 0.12$ & $^{+0.16}_{-0.14}$
	& $^{+35}_{-34}$ & $^{+170}_{-120}$ \\
192 & 300 & $\pm 0.20$ & $\pm 0.16$ & $^{+0.21}_{-0.18}$
	& $^{+46}_{-44}$ & $^{+240}_{-150}$ \\
205 & 500 & $\pm 0.14$ & $\pm 0.11$ & $^{+0.12}_{-0.11}$
	& $^{+31}_{-29}$ & $^{+120}_{-100}$ \\
205 & 300 & $\pm 0.18$ & $\pm 0.13$ & $^{+0.17}_{-0.14}$
	& $^{+40}_{-38}$ & $^{+190}_{-110}$ \\
220 & 500 & $^{+0.10}_{-0.08}$ & $\pm 0.09$ & $^{+0.10}_{-0.08}$
	& $^{+28}_{-26}$ & $^{+100}_{-60}$ \\
220 & 300 & $^{+0.13}_{-0.12}$ & $\pm 0.12$ & $^{+0.13}_{-0.10}$
	& $^{+36}_{-33}$ & $^{+135}_{-90}$ \\
\end{tabular}
\end{table}


\begin{references}
\bibitem{lep}
	D. Schaile, {\sl Proceedings of the XXVII Int. Conf. on Highe
	Energy Physics}, eds. P.J. Bussey, I.G. Knowles,
	Glasgow, UK, 20-27 July 1994 (Inst. of Physics Publishing,
	1995) p. 27.
\bibitem{tgvreviews}
	For a recent comprehensive review of trilinear gauge boson couplings
	see: H. Aihara {\it et al.}, To appear in {\sl Electroweak
	Symmetry Breaking and Beyond the Standard Model}, eds. T. Barklow,
	S. Dawson, H. Haber and J. Siegrist (World Scientific),
	{\tt hep-ph/9503425}.
\bibitem{burgess94}
	C.P. Burgess, S. Godfrey, H. K\"onig, D. London, and I. Maksymyk,
	Phys. Rev. {\bf D49},6115 (1994).
\bibitem{dawson95}
	S. Dawson and G. Valencia, Nucl. Phys. {\bf B439}, 3 (1995).
\bibitem{hagiwara93}
	K. Hagiwara, S. Ishihara, R. Szalapski, and D. Zeppenfeld,
	Phys. Lett. {\bf B283}, 353 (1992); Phys. Rev. {\bf D48}, 2182 (1993).
\bibitem{tevatron}
	F. Abe {\it et al.} (CDF Collaboration), Phys. Rev. Lett.
	{\bf 74}, 1936 (1995);
	J. Ellison {\it et al.} (D0 Collaboration), Proceedings of the
	{\sl DPF'94 Conference}, Albuquerque, NM, August 1994;
	H. Aihara, to appear in the Proceedings of the
	{\sl International Symposium on Vector Boson
	Self-Interactions}, UCLA, February 1995;
	T.A. Fuess (CDF Collaboration), Proceedings of the
	{\sl DPF'94 Conference}, Albuquerque, NM, August 1994;
	F. Abe {\it et al.}, (CDF Collaboration), Proceedings of the
	{\sl 27th International Conference on High Energy Physics},
	Glasgow, Scotland, July 20-27, 1994;
	F. Abe {\it et al.}, (CDF Collaboration),
	FERMILAB-Pub-95/036-E, submitted to Phys. Rev. Lett.;
	S. Abachi {\it et al.}, (D0 Collaboration),
	FERMILAB-Pub-95/101-E, {\tt hep-ex/9505007}.
\bibitem{hagiwara87}
	K. Hagiwara, R.D. Peccei, D. Zeppenfeld, and K. Hikasa,
	Nucl. Phys. {\bf B282}, 253 (1987).
\bibitem{lep200}
	D. Zeppenfeld, Phys. Let. {\bf 183B}, 380 (1987);
	D. Treille {\it et al}, Proceedings of the ECFA Workshop on LEP 200,
	ed. A. B\"ohm and W. Hoogland, Aachen (1986), CERN 87-08, vol.2, p.414.
	D.A. Dicus, K. Kallianpur, Phys. Rev. {\bf D32}, 35 (1985);
	M.J. Duncan, G.L. Kane, Phys. Rev. Lett. {\bf 55}, 773 (1985);
        E.N.Argyres\ and C.G.Papadopoulos, Phys. Lett. {\bf B263}, 298(1991);
	G.Kane, J. Vidal, C.P. Yuan, Phys. Rev. {\bf D39}, 2617 (1990).
\bibitem{gintner95}
	M. Gintner, S. Godfrey, and G. Couture,
	Phys. Rev. {\bf D52} (in press).
%	Carleton University report OCIP/C 95-3.
\bibitem{sekulin}
	R.L. Sekulin, Phys. Lett. {\bf B338}, 369 (1994);
\bibitem{papa94}
	C.G. Papadopoulos, Phys. Lett. {\bf B352}, 144 (1995)
	({\tt hep-ph/9503276}).
\bibitem{berends95}
	F.A. Berends and A.I. van Sighem, INLO-PUB-7-95,
	{\tt hep-ph/9506391}.
\bibitem{couture92}
	G. Couture, S. Godfrey, and R. Lewis, Phys. Rev. {\bf D45},
	777 (1992);
	G. Couture and S. Godfrey, Phys. Rev. {\bf D49}, 5709(1994);
	P. Kalyniak, P. Madsen, N. Sinha, and R. Sinha,
	Phys. Rev. {\bf D48}, 5081 (1993).
\bibitem{beenakker94}
	For a detailed review see W. Beenakker and A. Denner,
	International Journal of Modern Physics {\bf A9}, 4837(1994).
\bibitem{barklow92}
	T. Barklow, {\sl Proceedings of the 1st Workshop on Physics
	with Linear Colliders}, Saariselka, Finland, Sept. 9-14 (1992).
\bibitem{fourfermions}
	F.A. Berends, R. Pittau, and R. Kleiss,
	Nucl. Phys. {\bf B424}, 308(1994);
	Nucl. Phys. {\bf B426}, 344(1994);
	G. Montagna, O. Nicrosini, G. Passarino, and F. Piccinini,
	Phys. Lett. {\bf B348}, 178 (1995);
	D. Bardin, A. Leike, and T. Riemann,
	CERN Report CERN-TH.7478/94 (1994, unpublished), {\tt hep-ph/9410361};
	J. Fujimoto {\it et al.}, KEP Report KEP 94-46;
	Y. Kurihara, D. Perret-Gallix, and Y. Shimizu,
	Phys. Lett. {\bf B349}, 367 (1995).
\bibitem{berends94b}
	F.A. Berends, R. Pittau, and R. Kleiss, Nucl. Phys.
	{\bf B426}, 344(1994).
\bibitem{beenakker91}
	W. Beenakker, K. Kolodziej, and T. Sack,
	Phys. Lett. {\bf B258}, 469 (1991);
	W. Beenakker and A. Denner, DESY 94-051.
\bibitem{fleischer94}
	J. Fleischer, F. Jegerlehner, K. Koodziej, and G.J. van Oldenborgh,
	preprint PSI-PR-94-16 (1994; unpublished);
\bibitem{emrc}
	O. Nicrosini and L. Trendadue, 	Nucl. Phys. {\bf B318}, 1 (1989);
	L. Trentadue {\it et al.}, in : Z Physics at LEP1, ed. G.
	Altarelli, CERN Yellow report CERN 89-08 (CERN, Geneva, 1989) p129;
	D. Bardin, M. Bilenky, A. Olchevski, and T. Riemann,
	Phys. Lett. {\bf B308}, 403 (1993);
\bibitem{altarelli}
	G. Altarelli, Proposed machine parameters for the LEP2
	physics study.
\bibitem{gaemers79}
	K. Gaemers and G. Gounaris, Z. Phys. {\bf C1}, 259 (1979).
\bibitem{cp}
	W.J. Marciano, A. Queijeiro, Phys. Rev. {\bf D 33}, 3449 (1986);
	F. Boudjema, K. Hagiwara, C. Hamzaoui, and K. Numata, Phys. Rev.
	{\bf D43}, 2223 (1991).
\bibitem{baur88}
	U. Baur and D. Zeppenfeld, Nucl. Phys. {\bf B308}, 127 (1988).
\bibitem{smloops}
	W.A. Bardeen, R. Gastmans, and B. Lautrup, Nucl. Phys.
	{\bf B46}, 319 (1982);
	K.J. Kim and Y.S. Tsai, Phys. Rev. {\bf D12}, 3972 (1975);
	E.N. Argyres {\it et al.,} Nucl. Phys. {\bf B391}, 23 (1993);
	J. Papavassiliou and K. Philoppides,
	Phys. Rev. {\bf D48}, 4255 (1993);
	G. Couture and J.N. Ng, Z. Phys. {\bf C35}, 65 (1987);
	G. Couture {\it et al.}, Phys. Rev. {\bf D38}, 860 (1988).
\bibitem{bagger93}
J. Bagger, S. Dawson, and G. Valencia, Nucl. Phys. {\bf B399}, 364 (1993).
\bibitem{nonlinear}
	B. Holdom, Phys. Lett. {\bf B258}, 156 (1991);
	A. Falk, M. Luke, and E. Simmons, Nucl. Phys. {\bf B365}, 523 (1991);
	T. Appelquist and G.H. Wu, Phys. Rev. {\bf D48}, 3235 (1993).
\bibitem{calkul}
	R.~Kleiss and W.~J.~Stirling, Nucl. Phys. {\bf B262}, 235 (1985);
	Z.Xu D.-H.Zhang L. Chang Nucl. Phys. {\bf B291}, 392 (1987).
\bibitem{monte}
	See for example V. Barger and R. Phillips, {\sl Collider Physics},
	(Addison-Wesley Publishing Company, 1987).
\end{references}
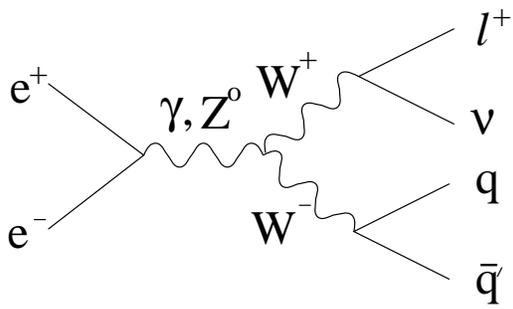
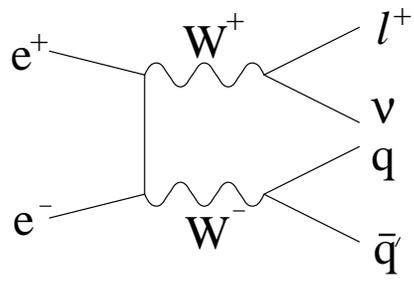
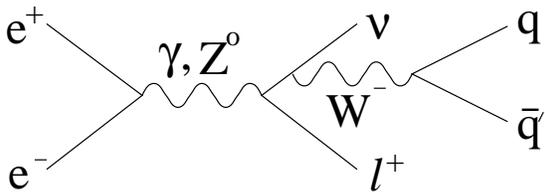
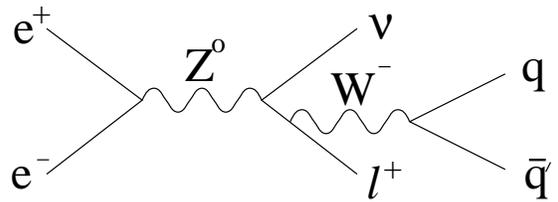
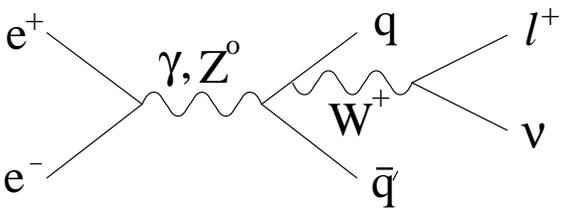
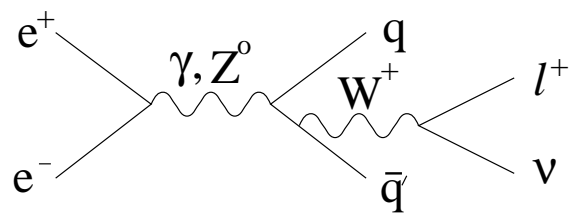
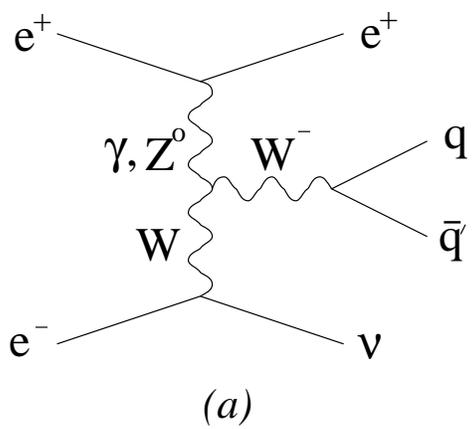
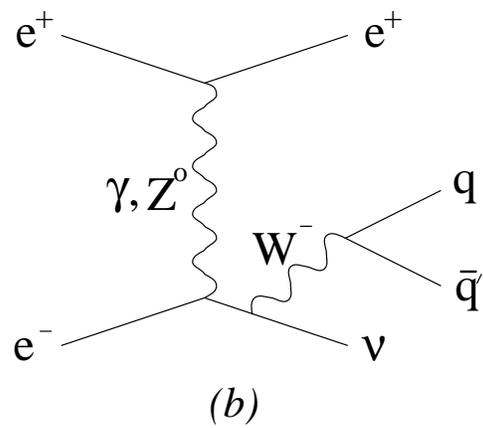
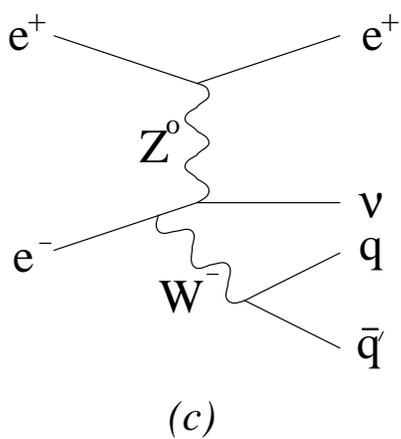
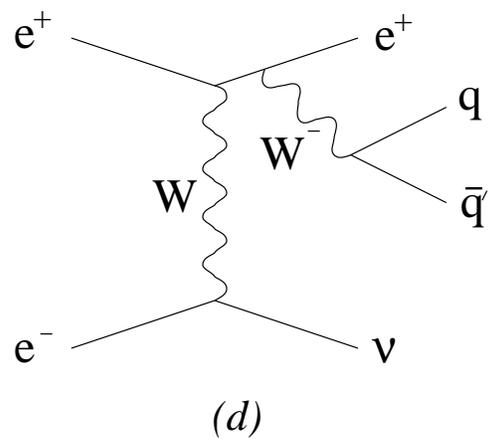
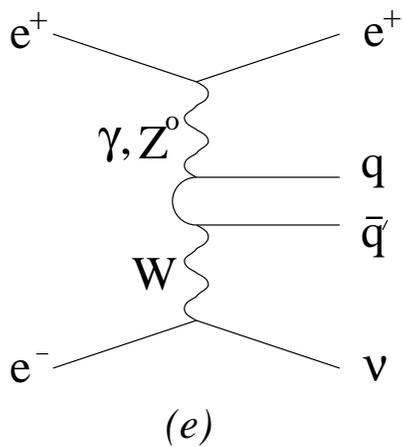
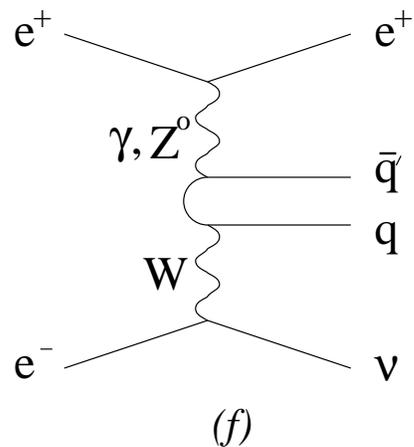
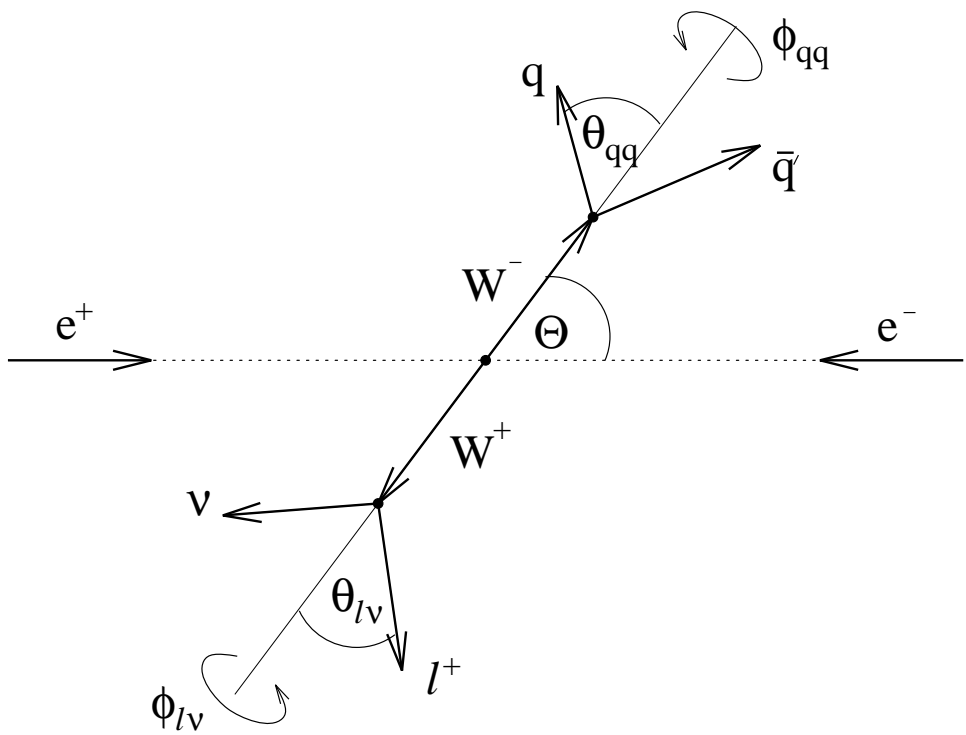
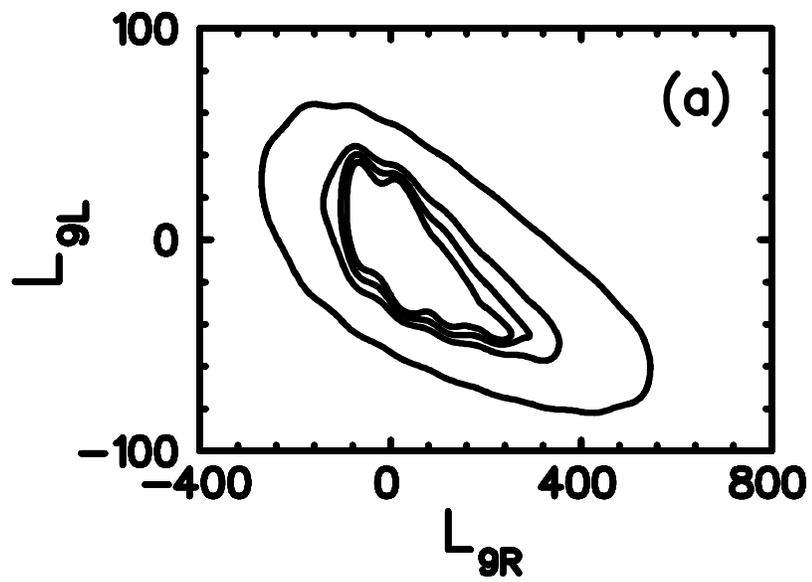
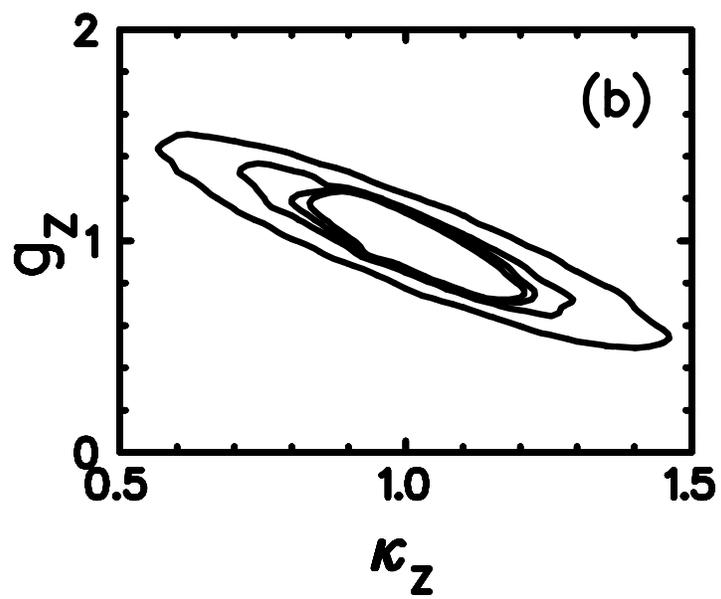
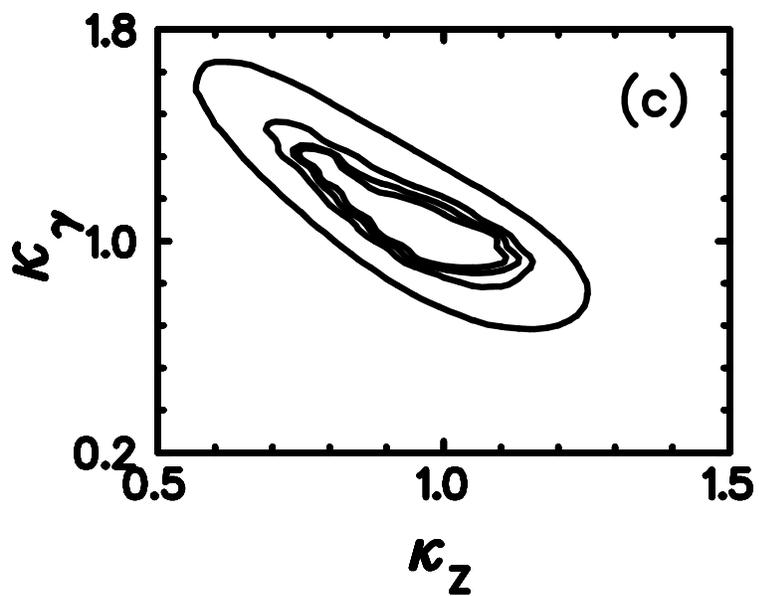
\end{document}